\documentclass[a4paper,fleqn,usenatbib]{mnras}
\usepackage{amssymb}
\usepackage{multirow}
\usepackage{morefloats}
\usepackage{amsmath}
\usepackage{bigstrut}
\usepackage{booktabs}
\usepackage{natbib}
\usepackage{float}
\usepackage{graphicx,epsfig,fancyhdr,epsf,txfonts,epstopdf}
\usepackage{latexsym,bbm}
\usepackage{lineno}
\usepackage{color}
\usepackage{ulem}

\title[Flares in the blazar W2R 1926+42]{Statistical analysis of variability properties of the Kepler blazar W2R 1926+42}

\author[Yutong Li et al.]{Yutong Li,$^{1,2}$\thanks{E-mail: qiuwutongyu622@163.com, liy@tcnj.edu}
 Shaoming Hu,$^{1}$\thanks{E-mail: husm@sdu.edu.cn}
 Paul J. Wiita,$^{2}$\thanks{E-mail: wiitap@tcnj.edu}
 and Alok C. Gupta$^{3,4}$
\\
\\
$^{1}$ Shandong Provincial Key Laboratory of Optical Astronomy and Solar-Terrestrial Environment, Institute of Space
Sciences, Shandong University, \\
~~Weihai, 264209, China\\
$^{2}$ Department of Physics, The College of New Jersey, 2000 Pennington Rd., Ewing, NJ 08628-0718, USA\\
$^{3}$ Aryabhatta Research Institute of Observational Sciences (ARIES), Manora Peak, Nainital 263 002, India\\
$^{4}$ Key Laboratory for Research in Galaxies and Cosmology, Shanghai Astronomical Observatory, Chinese Academy of Sciences, 80 Nandan Road, \\ Shanghai 200030, China
}

\date{Accepted XXX. Received YYY; in original form ZZZ}

\pubyear{2018}

\begin{document}
\label{firstpage}
\pagerange{\pageref{firstpage}--\pageref{lastpage}}
\maketitle

\begin{abstract}
   \noindent
  We analyzed Kepler light curves of the blazar W2R 1926$+$42 that provided nearly continuous coverage from quarter 11 through quarter 17 (589 days between 2011 and 2013) and examined some of their flux variability properties. We investigate the possibility that the light curve is dominated by a large number of individual flares and adopt exponential rise and decay models to investigate the symmetry properties of flares. We found that those variations of W2R 1926+42 are predominantly asymmetric with weak tendencies toward positive asymmetry (rapid rise and slow decay).  The durations ($D$) and the amplitudes ($F_{0}$) of flares can be fit with log-normal distributions. The energy ($E$) of each flare is also estimated for the first time. There are positive correlations between log$D$ and log$E$ with a slope of 1.36, and between log$F_{0}$ and log$E$ with a slope of 1.12. Lomb-Scargle periodograms are used to estimate the power spectral density (PSD) shape. It is well described by a power law with an  index ranging between $-1.1$ and $-1.5$. The sizes of the emission regions, $R$, are estimated to be in the range of 1.1 $\times$ 10$^{15}$cm -- 6.6 $\times$ 10$^{16}$cm. The flare asymmetry is difficult to explain by a light travel time effect but  may be caused by differences between the timescales for acceleration and dissipation of high-energy particles in the relativistic jet.  A jet-in-jet model also could produce the observed log-normal distributions.
\end{abstract}

\begin{keywords}
galaxies: active -- BL Lacertae objects: individual: W2R 1926+42 -- techniques: photometric
\end{keywords}



\section{Introduction}

  Blazars are the most active class of active galactic nuclei (AGNs). They have relativistic charged particle jets which are nearly aligned to the observer's line of sight \citep{Bland79}. Blazars are classified into two main subclasses, namely BL Lacertae objects (BL Lacs) and flat spectrum radio quasars (FSRQs). BL Lacs either do not show, or show extremely weak, emission lines whereas FSRQs show strong emission lines in their optical spectra \citep{Urry95}. Blazars are characterized by their rapid flux variability across the complete electromagnetic (EM) spectrum, large and variable polarization from radio to optical bands, and emission that is predominantly nonthermal. The spectral energy distributions (SEDs) of blazars are characterized by two distinct and broad peaks located respectively in the NIR to X-ray and in the $\gamma-$ray regimes \citep{Ulr97,Fos98}. The former peak is unambiguously identified as synchrotron radiation from ultrarelativistic electrons. The origin of the second component, sometimes extending to TeV energies, is somewhat less clear. In the more popular leptonic models, the emission originates from the inverse Compton (IC) scattering of the ambient photons by the same electrons that produce the synchrotron emission \citep{Mara92, Bot07}. The seed photons can be synchrotron photons in the synchrotron-self-Compton (SSC) models or be external to the jet, i.e., external Compton (EC) models. On the other hand, in hadronic models, processes such as proton synchrotron and radiation produced by secondary particles are suggested to explain the emission of high energy to very high energy $\gamma -$ray radiation from blazars \citep{Man93}. Based on the peak frequency ($\nu^{sync}_{p}$) of the synchrotron component, blazars frequently are divided into three sub-classes: low-synchrotron peaked (LSP: $\nu^{sync}_{p} < 10^{14}$ Hz), intermediate synchrotron peaked (ISP: $10^{14} < \nu^{sync}_{p} < 10^{15}$ Hz) and high-synchrotron peaked (HSP: $\nu^{sync}_{p}> 10^{15}$ Hz) blazars \citep{Abdo10a}.

  Rapid and large-amplitude variability is a significant property of  blazars and it occurs over the entire range of observable timescales from minutes to decades. The bulk of this variability, particularly the largest changes, is thought to arise from shocks propagating down the jet \citep{Mars85, Mars08}, and provides a useful tool to diagnose the physical mechanisms responsible for the emission of blazars. Intrinsic brightness variations of blazars could be affected by a variety of physical conditions, such as the size of the emission region, changes in densities of radiating plasma, or the strength and direction of magnetic fields, while more extrinsic changes arise from the change in velocity, and hence Doppler factor, of those regions. The observability of emission from a blazar jet also may be dependent on the opacity of the emission region at different wavelengths. Systematic investigations of flux variability provide crucial information about the location of the dominant emission region in the jet \citep{Mars10}.

 There have been previous investigations into the temporal symmetry (or asymmetry) of blazar variations, which is the focus of our current study. Generally symmetric profiles of flares at different frequencies have been found by \cite{Abdo10b}. They analyzed ten $\gamma$-ray blazars and found that most of the flares had symmetric profiles. Also \cite{Chat12} presented the light curves of six blazars and showed that the prominent flares  in both optical-IR and $\gamma-$ray light curves are predominantly symmetric. The phenomenon that most flares at different timescales present symmetric profiles may stem from a light travel time effect. Due to the existence of a finite size and the resultant finite duration of the evolution of an internal shock, processes occurring faster than the light-travel time through the region size or the finite duration of the event will be smoothed out \citep{Chiab99,Kat00,Sok04,Chen11}. Therefore, the observations can be affected by the light-travel time. Nonetheless, asymmetric flares exist \citep{Val99,Saito13,Wie16,Li17}. \cite{Val99} presented the long-term light curves of blazars at 22 and 37 GHz and found that the sources had asymmetric profiles with the mean ratio of the decay to rise timescales being $\sim$1.3. \cite{Saito13} reported that some flares from the blazar PKS 1510$-$089 exhibited asymmetric profiles. \cite{Wie16}   studied the hard X-ray observations of the blazar S5 0716+714 and also showed that all flares in optical, ultraviolet and $\gamma-$ray bands in 2015 January and February had asymmetric profiles. Optical light curves of the same source using about 20 years observations have  analyzed by \cite{Li17} who found that flares on both intraday and short-term (days to weeks) timescales had predominantly asymmetric profiles, with decay timescales $\sim$1.3  times longer than rise timescales.

 Several models have been proposed to understand the acceleration mechanisms and behavior of the shocked plasma, and to our knowledge, two of them are capable of simultaneously reproducing multi-wavelength light curves, time-dependent SEDs, and multi-frequency polarization. The first of these is the turbulent extreme multi-zone model (TEMZ)  developed by \cite{Mars14}. It assumes that the jet geometry includes a dominant region which includes a turbulent chaotic magnetic field which can be described in terms of many cylindrical computational cells, and where each cell possesses an uniform magnetic field whose direction is selected randomly. A proper integration of the time-dependent contributions from individual cells to yield the complete emission is performed to simulate the observations. The second method, known as the helical magnetic field model (HMFM) \citep{Zhang14}, assumes an ordered helical magnetic field and takes into account all light travel time effects (LTTEs) employing a Monte-Carlo / Fokker-Planck (MCFP) code to recreate the SEDs, light curves, and polarization. However, neither model can naturally explain the symmetric flaring activities.

 The blazar W2R 1926+42 is classified as a LSP BL Lac object at a redshift $z=0.154$ \citep{Edel12} that happened to fall in the original field studied by the {\it Kepler} satellite. Because of the limitations of ground-based optical monitoring, very rarely could continuous coverage of a blazar be obtained over more than a single 12 hr night, even when multiple telescopes were employed. While the {\it Kepler} satellite, with its nearly continuous coverage for almost three months on a single module, beautifully overcomes that limitation, that field contained very few previously known AGN \citep{Mush11,Weh13} and none of the known FSRQs was particularly active during the original mission \cite{Rev14}.  However, other AGN were subsequently discovered in the field \citep{Edel12}. The blazar W2R 1926+42 was found to be highly variable and so some of its properties have been studied before \citep{Edel13,Bachev15,Mohan16,Sasada17}. \cite{Edel13} reported {\it Kepler} light curves in Quarter 11 (Q 11) and Q 12 with 30-minute time sampling, and showed that the flux distribution is highly skewed and non-Gaussian. \cite{Bachev15} searched for the signature of low-dimensional chaos in the light curve of the blazar by applying a correlation integral method to both real datasets and phase randomized surrogates. They did not find low-dimensional chaos in the light curve. \cite{Mohan16} studied the long term {\it Kepler} light curve from Q11 to Q17 and showed that the normalized excess variance $F_{var}$ ranged from 1.8 percent in the quiescent phase to 43.3 percent in the outburst phase, and obtained a black hole mass of $M_{\bullet} = (1.5 - 5.9)\times 10^{7} M_{\odot} $ using $F_{var}$. They performed time series analyses of the full interpolated light curve as well as of six segments (each segment including $\sim$100 d) to study the power spectral density (PSD) shape. They found a weighted mean PSD slope of -1.5 $\pm$ 0.2, and inferred the presence of a possible quasi-periodic oscillation (QPO) peaking at 9.1 d using wavelet analysis methods; however, this is quite uncertain as it only lasted 3.4 cycles. Using the  \textquotedblleft shot analysis\textquotedblright  ~technique, \cite{Sasada17} analyzed the Kepler light curve of Q14 with 1-minute integrations. An averaged profile of 195 selected shots showed that it was composed of a spiky-shape (exponential rise and decay) shot component and two slow varying components around the peak time. In order to systematically investigate the variety of the time profiles of flux variability in individual flares, here we present our analysis of the full {\it Kepler} light curves of the blazar W2R 1926+42, running from Q11 through Q17 (589 days).

 This paper is organized as follows. Section 2 presents the data reduction and Section 3 gives the results of various time-series analyses using these data. Section 4 provides a discussion of the implications of these results and our conclusions are given in Section 5.

\section{Data and Light curves}

 {\it Kepler} monitored a $\sim 115$ deg$^{2}$ region of the sky, sampling pixel maps every $\sim$30 minutes with a $> 90\%$ duty cycle, thus producing light curves covering several years \citep{Bor10,Koch10}. {\it Kepler} monitoring of W2R 1926+42 began in Q11. The {\it Kepler} data applied in this paper are from Q11 to Q17, spanning the 589 days from 2011 September 29 to 2013 May 11, with a median sampling rate of 0.02d, and containing only small data gaps. A list of the temporal intervals associated with each quarter of data is given in Table~\ref{tab:quartertime}. The light curve is extracted using the standard {\it Kepler} procedure (Simple Aperture Photometry (SAP) Flux Data after the Pre-search Data Conditioning module (PDCSAP, or just PDC, Flux Data), covering roughly the $4300-8900 {\AA}$ range) which is automatically applied to every object and is available on the Kepler data archives\footnote{\url{http://archive.stsci.edu/kepler/}}. A detailed description of the light curve extraction has been presented in \cite{Bachev15}. The original version of the PDC Flux Data from {\it Kepler} was designed to prioritize searches for exoplanet transits and had a significant deficiency for examining AGN data in that astrophysical features were often removed as a side-effect to the removal of instrumental effects. A new PDC version, which utilizes a Bayesian approach for removal of systematics, also reliably corrects most instrumental artifacts in the light curves while at the same time preserving planet transits as well as other astrophysically interesting signals. This new version is currently only used for {\it Kepler} long-cadence light curves but not for short-cadence light curves \citep{Stumpe12}. In this paper, we present our analysis using this improved long-cadence PDC Flux data (corrected flux time series); however, we also make a brief comparison using (SAP) Flux data (uncorrected flux time series for simple aperture photometry).

  \begin{table*}
   \centering
   \caption{Kepler Mission Quarter Dates}
   \label{tab:quartertime}
   \begin{tabular}{lccccccc}
   \hline\hline
   Name                & Q11         & Q12         & Q13         & Q14         & Q15         & Q16         & Q17 \\\hline
   Beginning Date  & 29-Sep-2011 & 05-Jan-2012 & 29-Mar-2012 & 28-Jun-2012 & 05-Oct-2012 & 12-Jan-2013 & 09-Apr-2013 \\\hline
   Ending Date   & 04-Jan-2012 & 28-Mar-2012 & 27-Jun-2012 & 03-Oct-2012 & 11-Jan-2013 & 08-Apr-2013 & 11-May-2013 \\\hline
   \end{tabular}
 \end{table*}

\begin{figure*}
    \centering
    \includegraphics [trim=0.2cm 3cm 0.2cm 3cm,width=0.8\textwidth,clip]{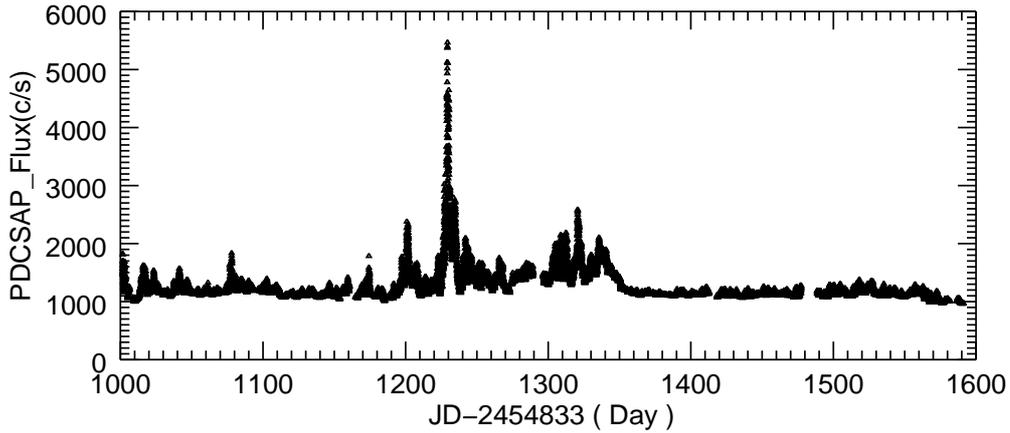}
    \caption{The long-term light curve of W2R 1926+42 from Kepler.}
    \label{fig:lightcurve}
 \end{figure*}

 The light curve of the object W2R 1926+42 obtained by Kepler is presented in Figure \ref{fig:lightcurve}. Strong activity over timescales from minutes to days is clearly seen. The light curve is composed of a period of modest variability in the first $\sim$180~d, with much larger variability activity for the next $\sim$190~d, which was followed by a relatively quiescent phase lasting the last $\sim$220~d.

 To analyze the variability properties of W2R 1926+42, our approach to the {\it Kepler} data was to fit them to a series of (usually overlapping) structures, or ``flares'' instead of  employing the more common assumption of a continuous process examined in terms of power-spectral analyses \citep{Weh13,Edel13,Rev14}.  As noted above, this type of decomposition has been used for several other active galactic light curves over a wide range of bands \citep{Val99,Abdo10b,Chat12,Saito13,Wie16,Li17}. Light curves calculated over shorter time intervals reveal more detail and potentially shorter flares. Our fits were done directly without assuming any {\it a priori} binning. In this method, some short flares may not have been reliably detected, but this should not affect the statistical results presented below. We separated the long-term light curves into small segments at the troughs of the light curves, so as to avoid severing  flares. The longest segments were not longer than 10 days, and the shortest segment was only about 1.5 days though most of the segments were about 7 -- 8 days long.

\begin{figure*}
    \centering
    \includegraphics [trim=0.2cm 0.2cm 0.2cm 0.2cm,width=0.8\textwidth,clip]{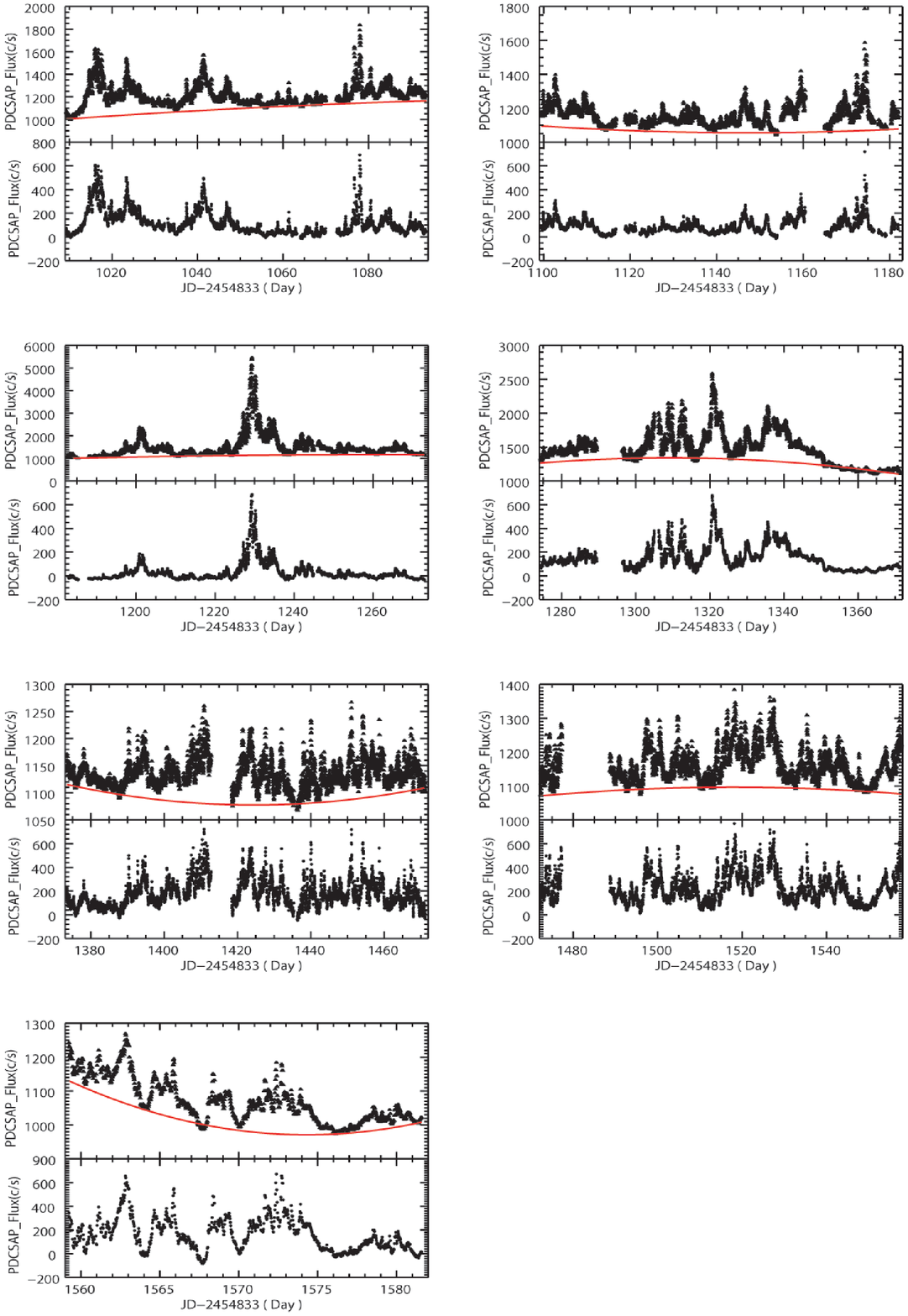}
    \caption{W2R 1926+42 light curves for Qs 11--17, respectively. Top panels show light curves
    with detected variability and polynomial functions approximating the long-term quarterly baseline component underlying
    the variability are shown as (red) solid curves. Bottom panels show the light curves after subtraction of these
    best-fit polynomial functions.}
    \label{fig:quarter}
 \end{figure*}

 A systematic analysis of all active variability revealed that the light curves indeed can reasonably be modeled as consisting of many flares. In order to remove the long-term variability trends before proceeding to fit the properties of these individual flares, we need to subtract baseline components.  These baselines were fit to polynomial curves  and then subtracted from the total fluxes in order to isolate the flaring components. Before performing these polynomial fits, each quarter was divided into several smaller segments, each of which contained $\sim$750 data points (though  Q17 provided only 250 data points, because of early termination of data taking caused by the loss of {\it Kepler's} second reaction wheel). Then the polynomial baselines in each corresponding quarter were fitted using the minimum values found during each segment employing the Interactive Data Language (IDL) routine POLYFIT. Figures \ref{fig:quarter} show the light curves from Q11 to Q17. The top panels show the revised PDC {\it Kepler} light curves of W2R 1926+42, where the solid curve is a polynomial function approximating a long-term baseline component, while the bottom panels show the light curves after subtraction of the best-fit polynomial function for each quarter. The bottom light curves of Figure \ref{fig:quarter} are adopted in this paper to examine the variability properties of W2R 1926+42.

\section{Time-series Analyses}

\subsection{Flare profile models}

 Blazars' light curves can be considered to be composed of overlapping flares caused by events in the jet and/or in the accretion disk / corona region \citep{Val99, Abdo10b, Chat12, Wie16, Guo16, Sasada17}. The statistical analysis of individual flares allows us to most  directly access the characteristic timescales involved in shaping the energy production and dissipation processes in the source. In this case, these can provide constraints on the locations and the structures of emission zones in the blazar W2R 1926+42.

\begin{figure*}
    \centering
    \includegraphics [trim=0.2cm 0.2cm 0.2cm 0.2cm,width=0.8\textwidth,clip]{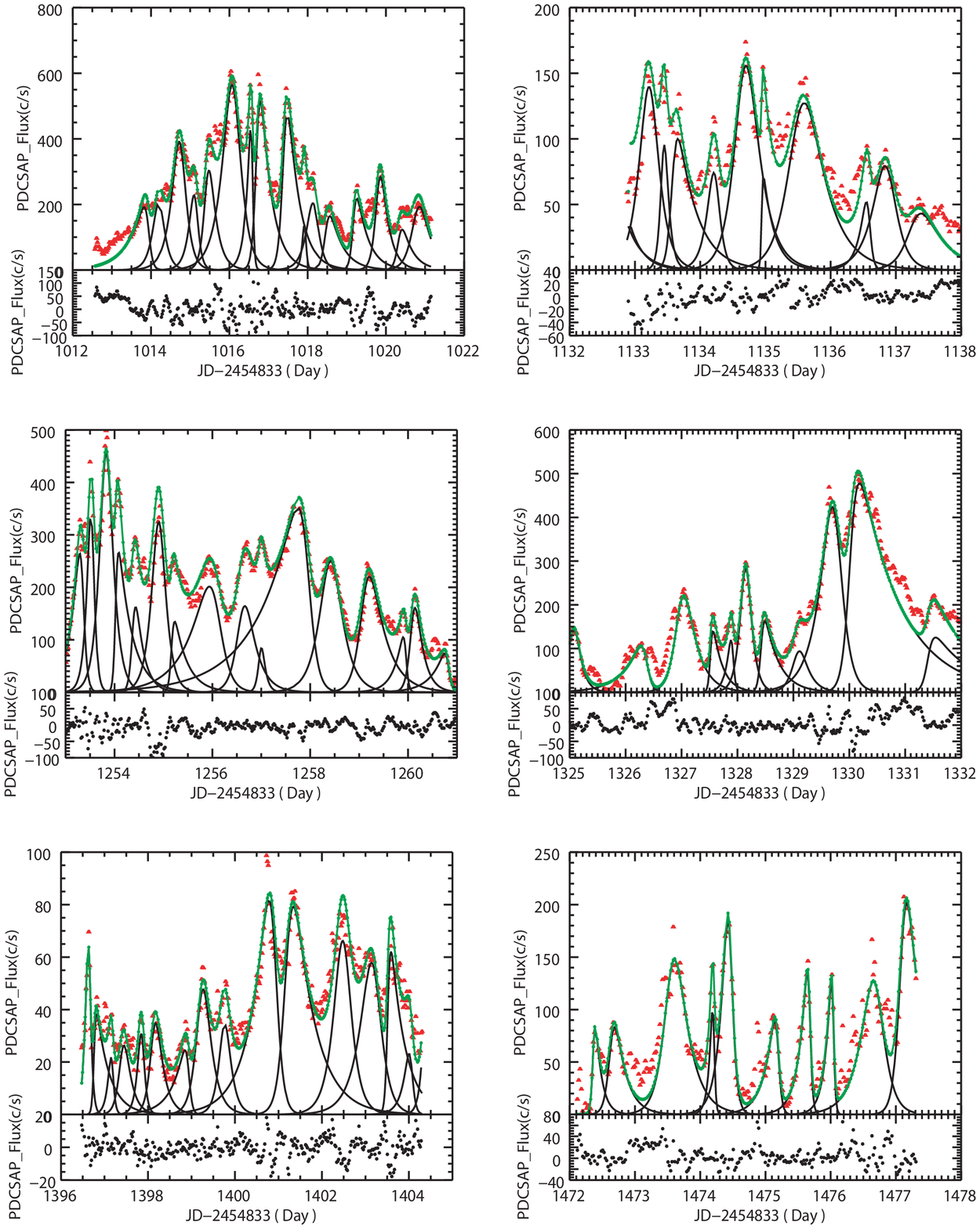}
    \caption{Examples for decomposition of W2R 1926+42 light curves. In each upper panel, red triangles
    denote the Kepler data, black solid lines show the fitted individual flares, green pluses and connecting lines correspond to
    superposition of all decomposed flares added to the constant background level and in each lower panel the black
    asterisks show the residual fluxes (data $-$ summed model).}
    \label{fig:flare}
 \end{figure*}

 We characterize the shape of a flare in terms of a function of exponential rise and decay proposed by \cite{Abdo10b}:
  \begin{equation}
   F(t)= F_{c} +  \frac{2 F_{0} }{{\rm exp} [(t_0-t)/T_{r}]+{\rm exp}[(t-t_0)/T_d]},
  \label{equ:fit}
  \end{equation}
 where $T_{r}$ and $T_{d}$ measure the timescales of the rise and decay phases, $F_{c}$ is a constant level underlying the flare, $F_{0}$ represents the amplitude of a flare and $t_{0}$ is the time of the peak of each flare. Because we subtract polynomial baseline components from the light curves we can take $F_{c}$ to be 0 in this paper. First, we fit the highest peak of each light curve segment to an exponential rise and decay and get values of $t_0, F_0, T_r$, and $T_d$. We then subtract this fitted flare from the light curve and fit the next highest remaining peak to another exponential rise and decay and then continue this process; a detailed description of our fitting method has been given in \cite{Li17}. The difference from \cite{Li17} involves the limiting condition on the minimum number of flares required to adequately model the light curves. In that work, the fitting of flares was repeated until the standard deviation change of residual flux was less than $1\%$ compared with the previously fitted flare. Due to the very dense sampling of {\it Kepler} data, the median time intervals of W2R 1926+42 (0.02 d) is much shorter than that for S5 0716+714 (0.6 d) used in \cite{Li17}, thus the standard deviation change of residual fluxes during fitting W2R 1926+42 flares will be much smaller. Many small flares that could be best considered as instrumental noise will be overly fitted if we used the same limiting condition as \cite{Li17}. In this work, we changed the limiting condition so that the amplitude of residual flux was not larger than $20\%$ of the amplitude of the highest flare in corresponding segment. As shown, e.g., in the top-right panel of Figure \ref{fig:flare}, some incomplete flares can also be fitted on the edges of segments. In order to analyze the properties of variability reasonably, we did not consider these incomplete flares.

  For each flare, we calculated the total flare duration as,
  \begin{equation}
  D = 2 (T_{r}+T_{d}),
  \label{equ:dur}
  \end{equation}
  which, for nearly symmetric flares, corresponds to the interval where the flux level on either side is at about 20\% of the peak value;
  this approach is also used in \cite{Abdo10b}.
  The time asymmetry, or  skewness, parameter $K$, as defined in \cite{Abdo10b}, as,
  \begin{equation}
  K=(T_{d}-T_{r})/(T_{r}+T_{d}).
  \label{equ:skew}
  \end{equation}
 We also calculated the ``energy'', $E$ of each flare by integrating its flux (or count rate) over its duration.

 Six examples of the light curves and their decomposition into flares are shown in Figure \ref{fig:flare}. In each panel, the constant background flux $F_{c}$ is constrained to 0 and shown by the horizontal  lines. Triangles denote the Kepler data, solid curves show the fitted individual flares, pluses correspond to superposition of all decomposed flares and the asterisks in the bottom panels give the residual fluxes. Because of the limitation on the number of flares needed to fit each light curve, there still is apparent variability in the residual flux. The positive residuals indicate that we may miss some small flares and we also occasionally overfit the data, leading to negative residuals, but neither of these will significantly affect the statistical results which are based on a large number of flares. We fit a total of 740 flares using Equation~\ref{equ:fit}.

\subsection{Flare asymmetry}

\begin{figure}
    \centering
     \includegraphics[trim=0.2cm 0cm 1.3cm 0cm,width=0.36\textwidth,clip]{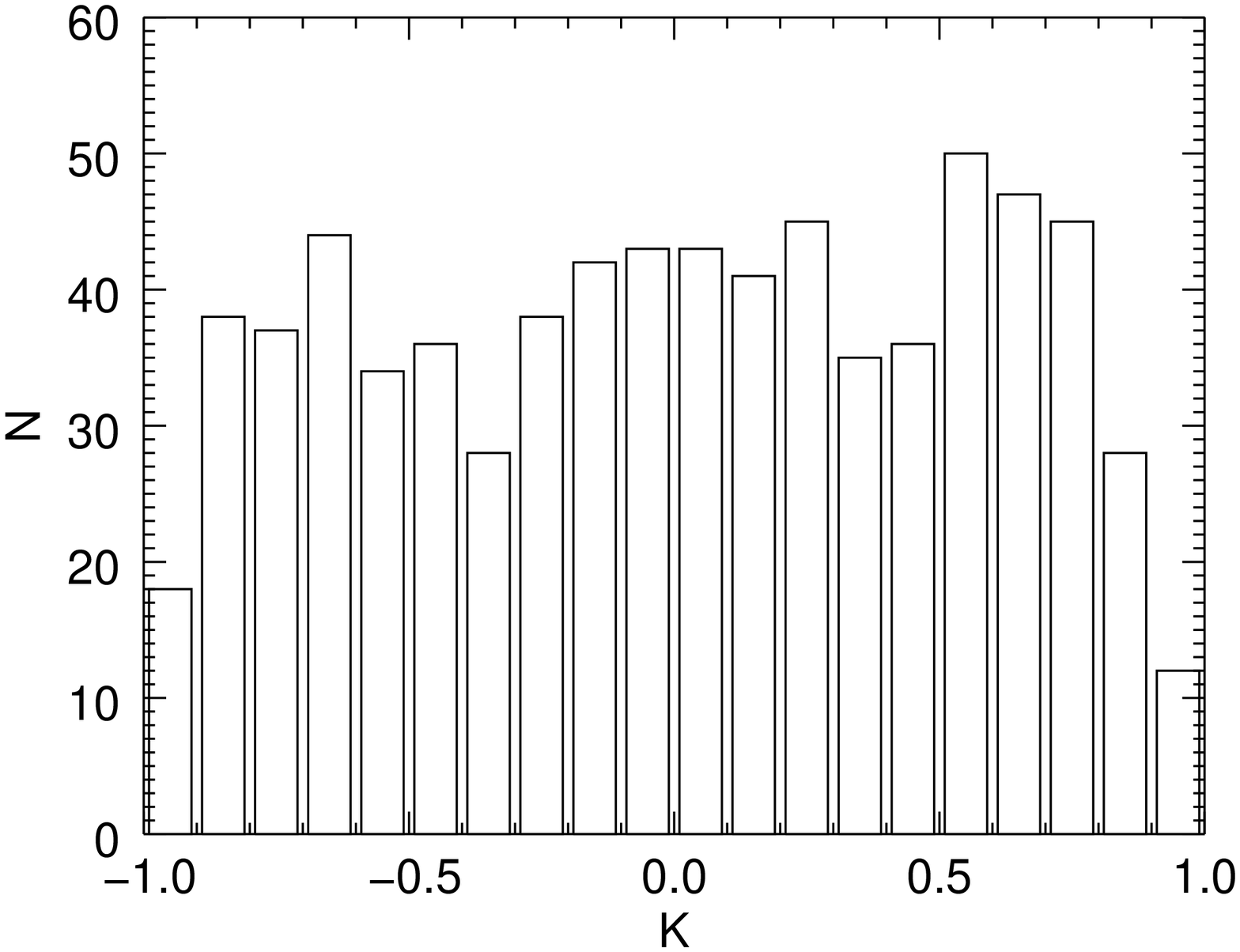}
     \caption{Distribution of the skewness parameter $K$ for all the decomposed light curves.}
     \label{fig:skewness}
 \end{figure}

  Study of the asymmetry parameter $K$ (Eq.~\ref{equ:skew}) gives insight into the flare profiles of W2R 1926+42. Histograms of the distribution of skewness parameters of these flares is shown in Figure \ref{fig:skewness}. The average value of $K$ is slightly positive at 0.05, but there is a wide distribution. The percentage of $K$ values in different ranges are listed in Table~\ref{tab:percent}. We see that the percentages of $K$ in these three different ranges are rather evenly distributed and note that the percentage of flares with $|K| > 0.3$  is greater than $60\%$, which indicates that the flares of W2R 1926+42 are predominantly asymmetric. Also, the percentage of $K > 0.3$ flares is larger than those for which $K < - 0.3$, providing evidence for at least a weak trend toward positive asymmetric profiles (more rapid rise and more gradual decay). This is in accordance with the nature of this phenomenon as seen for the blazar S5 0716+714  \citep{Hu14,Li17}.

 \begin{table}
   \centering
   \caption{Distributions of skewness parameters.}
   \space
   \label{tab:percent}
   \begin{tabular}{@{}cccccc@{}}\hline\hline
   Flare Num.&$|K|< 0.3$ & $< - 0.3$ & $> 0.3$ &Median\\
   \hline
   740       &34.0\%     & 31.8\%    &34.2\%   &0.05\\
   \hline
   \label{tab:percent}
   \end{tabular}
\end{table}

\subsection{Flare duration and flux variance}

\begin{figure*}
   \begin{minipage}{\textwidth}
   \centering
   \includegraphics[trim=0.2cm 0cm 1.3cm 0cm,width=0.36\textwidth,clip]{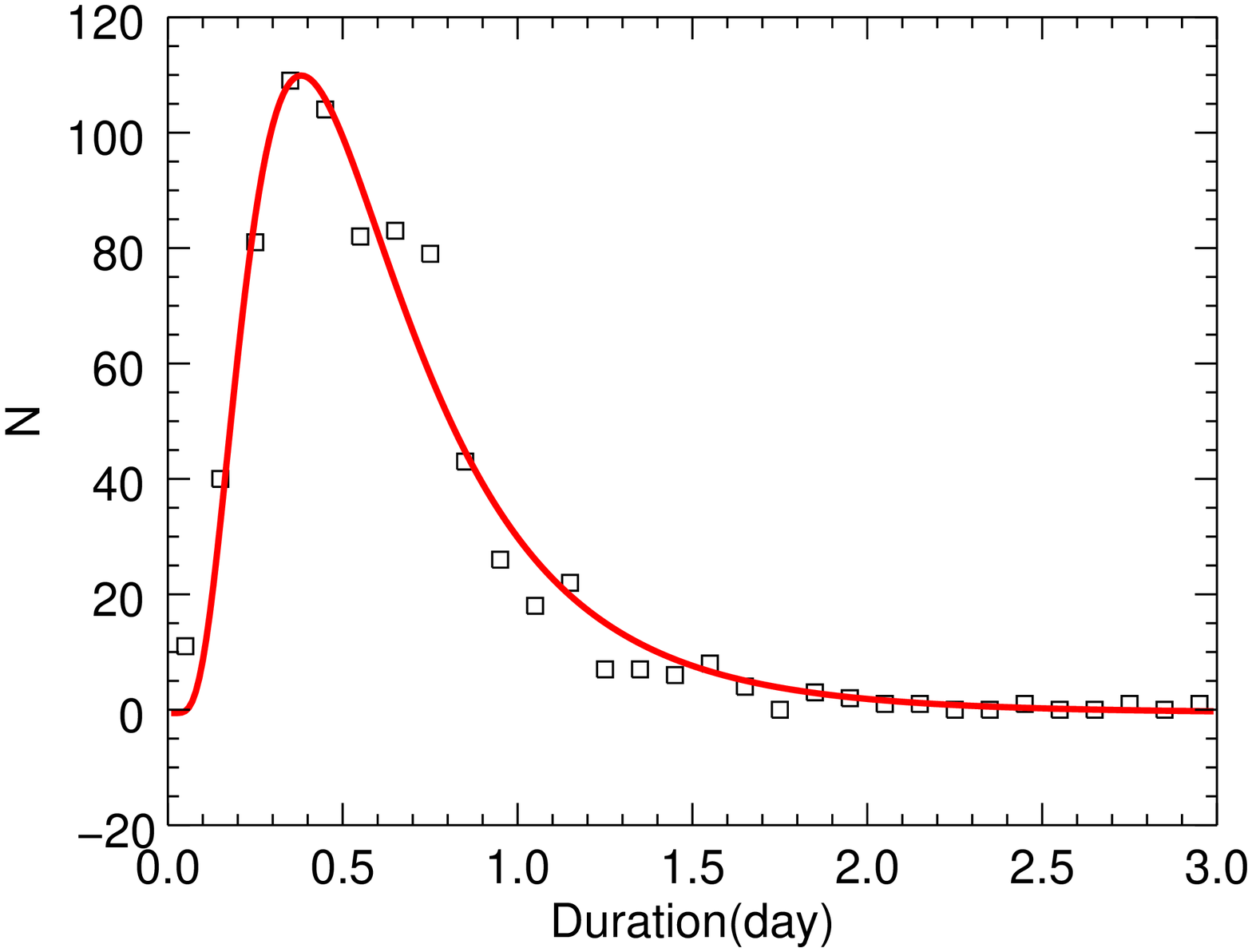}
   \includegraphics[trim=0.2cm 0cm 1.3cm 0cm,width=0.36\textwidth,clip]{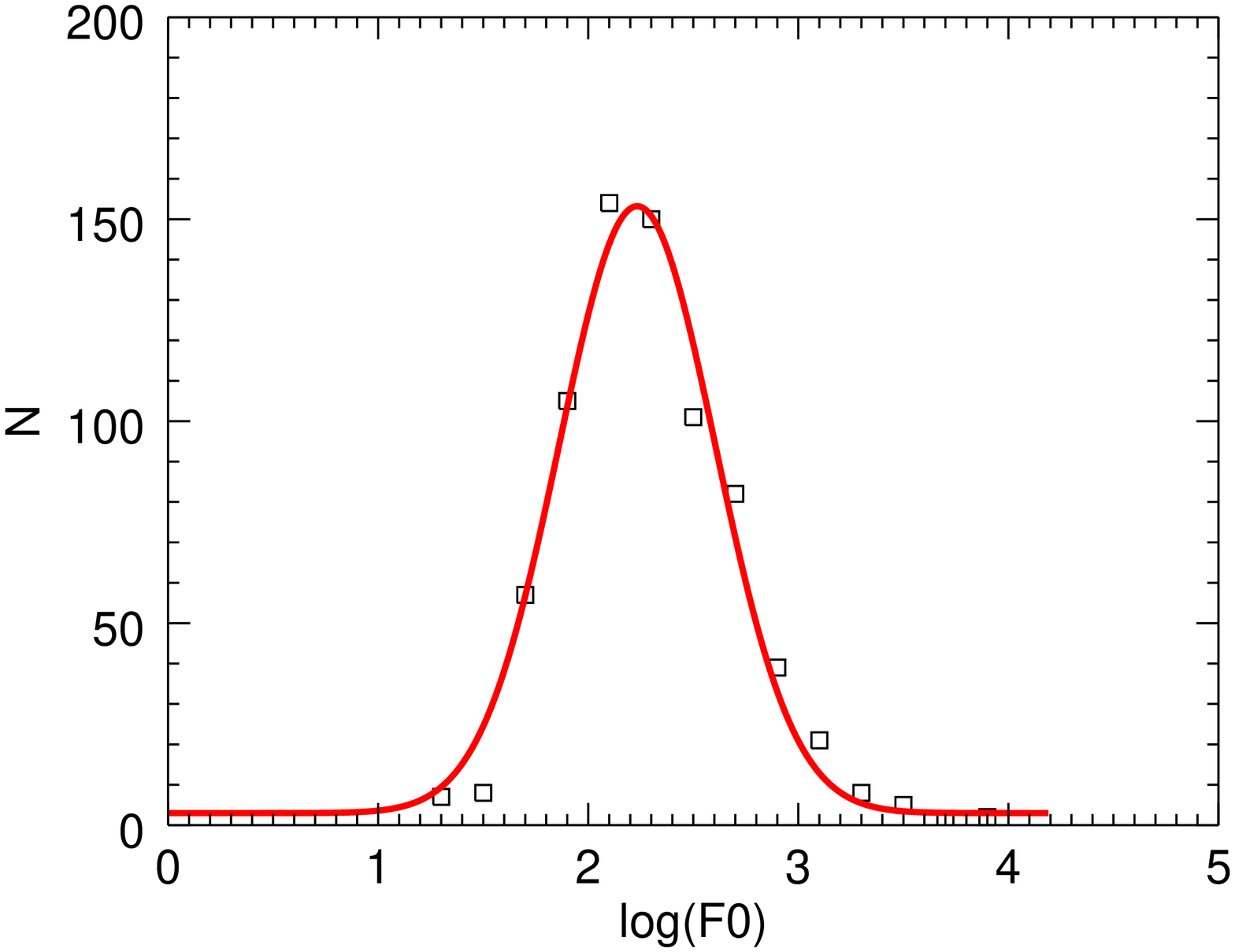}
   \end{minipage}\vspace{0.001cm}
   \caption{(left) Log-normal distributions of $D$ and (right) Gaussian distributions of log$(F_{0})$.}
   \label{fig:distribution}
 \end{figure*}

 In order to investigate the temporal properties of flares, we plotted the distributions of flare duration, $D$ (Eq.~\ref{equ:dur}), and flare flux amplitude, $F_{0}$,  in Figure \ref{fig:distribution}. We saw that the flare durations span the range of 0.05 to 3.0 d. The median of the flare durations is 0.54 d. The distribution of $D$ can be rather well represented by the log-normal function:
 \begin{equation}
  f(x)= \frac{1}{\sqrt{2\pi}  \sigma  x }  ~{\rm exp} \Bigl[{-\frac{(\ln x-\ln\mu)^2}{2 \sigma^2}}\Bigr],
 \end{equation}
 where $\mu$ and $\sigma^{2}$ respectively are the the mean and variance of the distribution \citep{Neg02}. The quality of the fits are quantified by the adjusted determination coefficient, $R^2$, where the closer the $R^2$ value is to 1, the better the fit. We find a value of $R^2 = 0.97$ to the form (4).

 We adopt a Gaussian on a platform to parameterize the logarithm of flux amplitude, log\ $F_{0}$:
  \begin{equation}
  f(x)= f_{0}+ \frac{A}{\sigma \sqrt{\pi / 2} } ~ {\rm exp} \Bigl[{-\frac{4( x-\mu)^2}{ \sigma^{2}}} \Bigr],
  \end{equation}
 where $f_{0}$ is a constant value, $A$ is the amplitude of the distribution and $\mu$ and $\sigma^{2}$  respectively correspond to the mean and variance of $x$ (log$F_{0}$). The fit is again excellent, with $R^2$ = 0.97. The results of these fits are summarized in Table~\ref{tab:logpara}.

 \begin{table*}
   \centering
   \caption{Model Fits and Correlations}
   \label{tab:logpara}
   \begin{tabular}{cccccccc}
   \hline\hline
   Distribution & parameter            &$R^2$ &   slope &$\mu$ & $\sigma$ &$f_{0}$ &$A$ \\\hline
   Log-normal   & $D$                  & 0.97 &   --    & 0.55 & 0.60     &  --    &  -- \\\hline
   Gaussian     & log$F_{0}$           & 0.97 &   --    & 2.23 & 0.74     &2.99    &139.98 \\\hline
   Linear       & log$D$ -- log$E$     & 0.47 &  1.36   &  --  &  --      & --     & --   \\
               & log$F_{0}$ -- log$E$ & 0.73 &  1.12   &  --  &  --      & --     & --   \\
   \hline
   \end{tabular}
 \end{table*}

  The log-normal distribution of duration $D$ and the Gaussian distribution of log $F_{0}$ are in accordance with the previous studies of W2R 1926+42 \citep{Edel13,Sasada17}. \cite{Edel13} showed the distribution of {\it Kepler} fluxes (only in Q11 and Q12) was clearly non-Gaussian, with a strong tail. The histogram of flux was fitted using a log-normal distribution. In our work, the profile of log $F_{0}$ shows an obvious Gaussian distribution which is essentially identical to that found for fewer data by  \cite{Edel13}. So, it is not surprising that the histogram of the measured fluxes were well fitted with log-normal distribution. This implies that the log-normal distribution situation is not caused by systematic error, but is a reflection of the origin mechanism. This type of distribution has been used in analyzing gamma-ray burst temporal structures \citep{Li96, Ehud02a, Ehud02b}, X-ray variability of BL Lac \citep{Gie09} and also previously in AGN optical variability work \citep{Guo16,Li17}. Evidence of log-normal variability had also been found in very high-energy $\gamma$-ray emission of the BL Lac PKS 2155-304 \citep{Deg08} and was taken to be the signature of an underlying multiplicative physical process.

\begin{figure*}
   \begin{minipage}{\textwidth}
   \centering
   \includegraphics[trim=0.2cm 0cm 1.3cm 0cm,width=0.36\textwidth,clip]{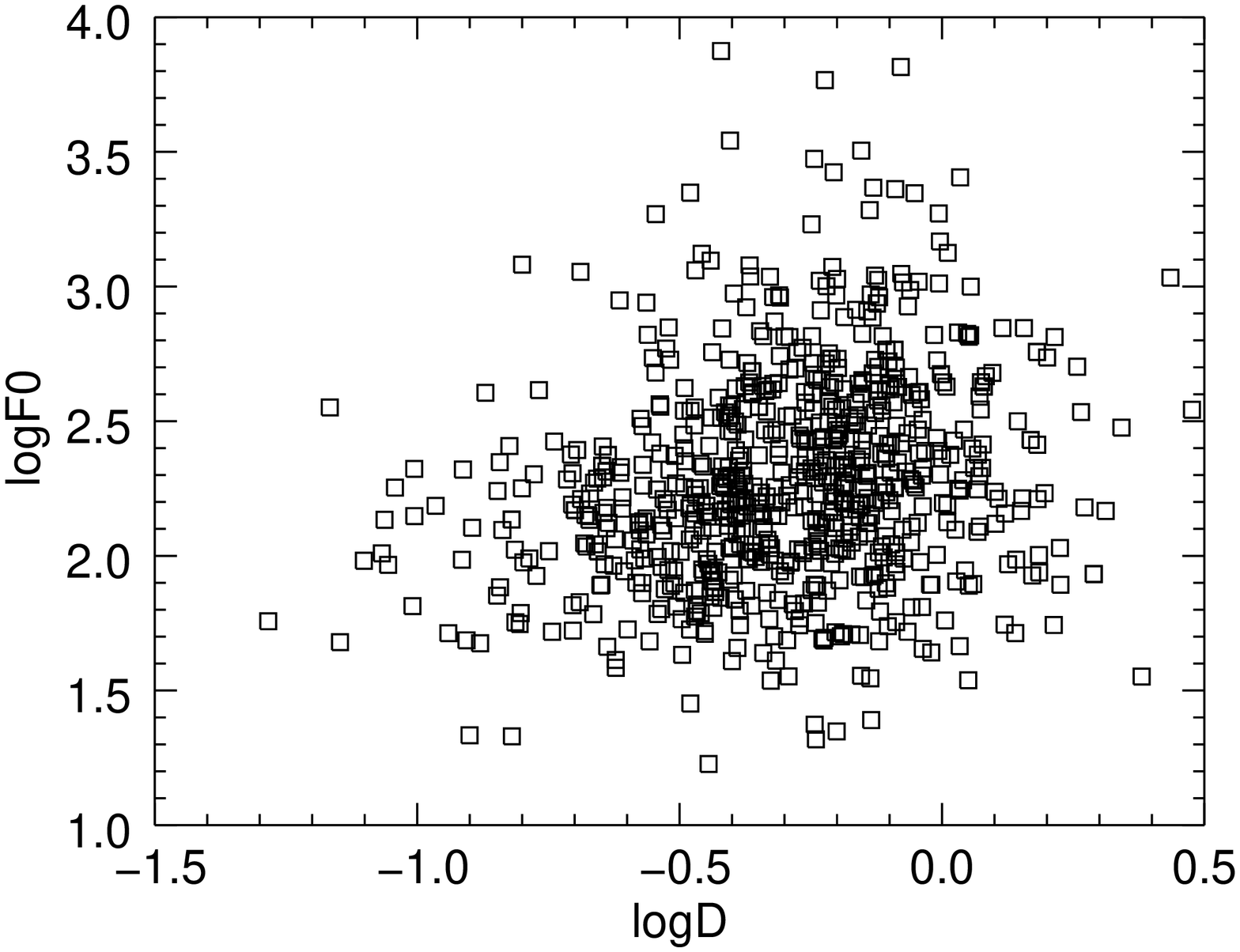}
   \includegraphics[trim=0.2cm 0cm 1.3cm 0cm,width=0.36\textwidth,clip]{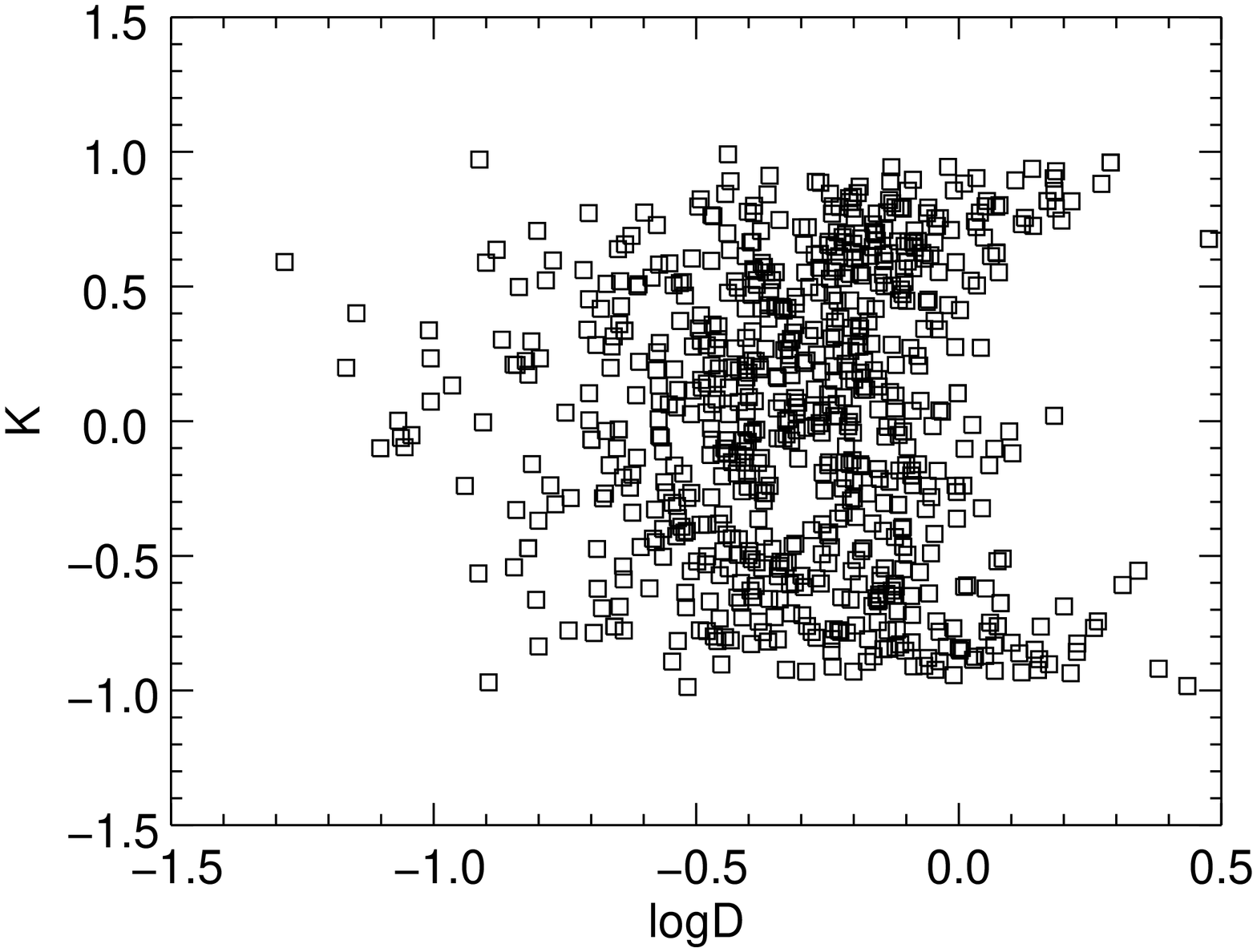}
   \end{minipage}\vspace{0.001cm}
   \caption{Possible correlations between flare durations and fluxes (left), and durations and skewness (right).}
   \label{fig:logDFV}
 \end{figure*}

 Plots of log$F_{0}$ and skewness $K$ of each flare against duration (log $D$) are shown in Figure \ref{fig:logDFV} but there is no evidence of any correlations between them.

\subsection{Flare energies}

\begin{figure*}
   \begin{minipage}{\textwidth}
   \centering
   \includegraphics[trim=0.2cm 0cm 1.3cm 0cm,width=0.3\textwidth,clip]{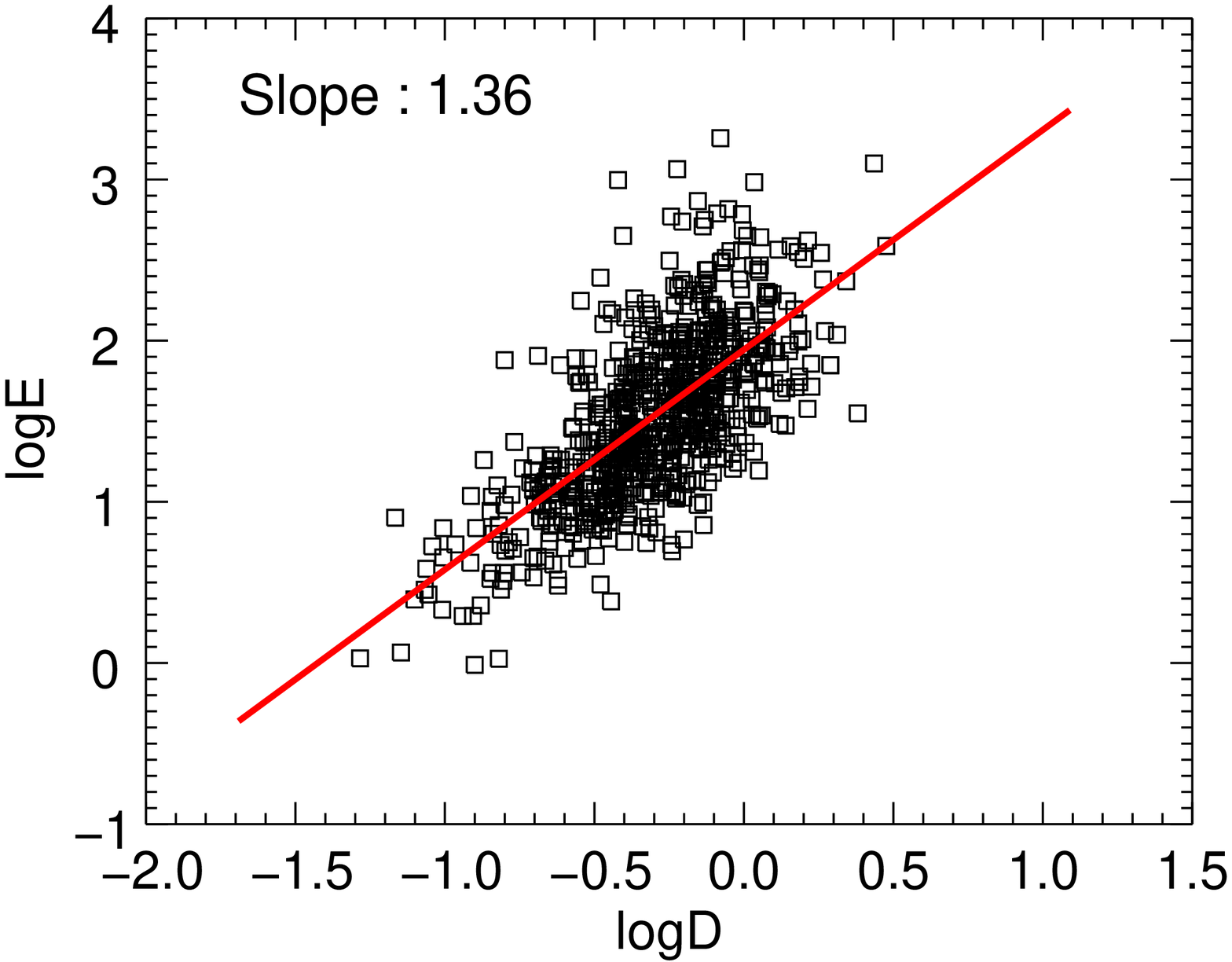}
   \includegraphics[trim=0.2cm 0cm 1.3cm 0cm,width=0.3\textwidth,clip]{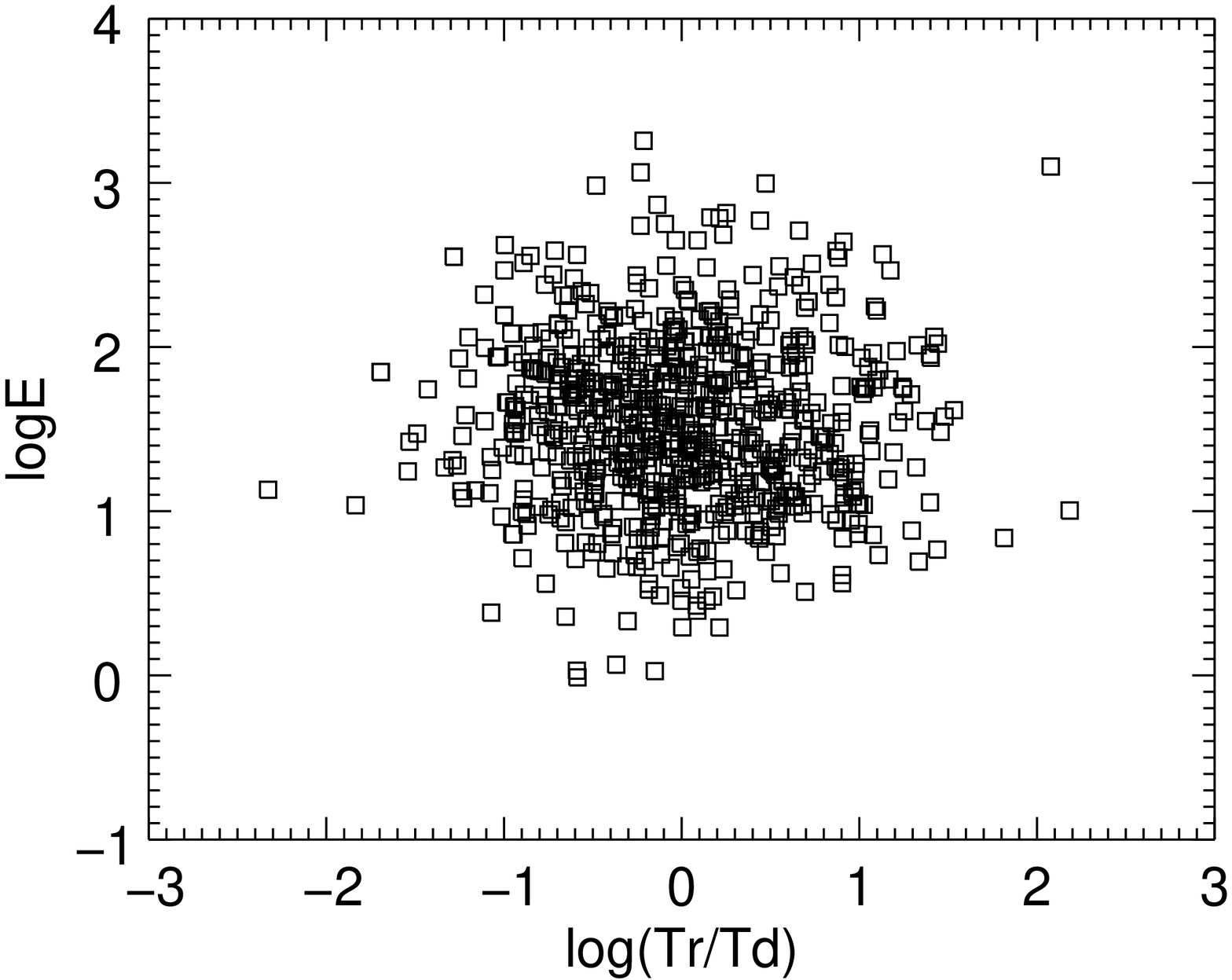}
   \includegraphics[trim=0.2cm 0cm 1.3cm 0cm,width=0.3\textwidth,clip]{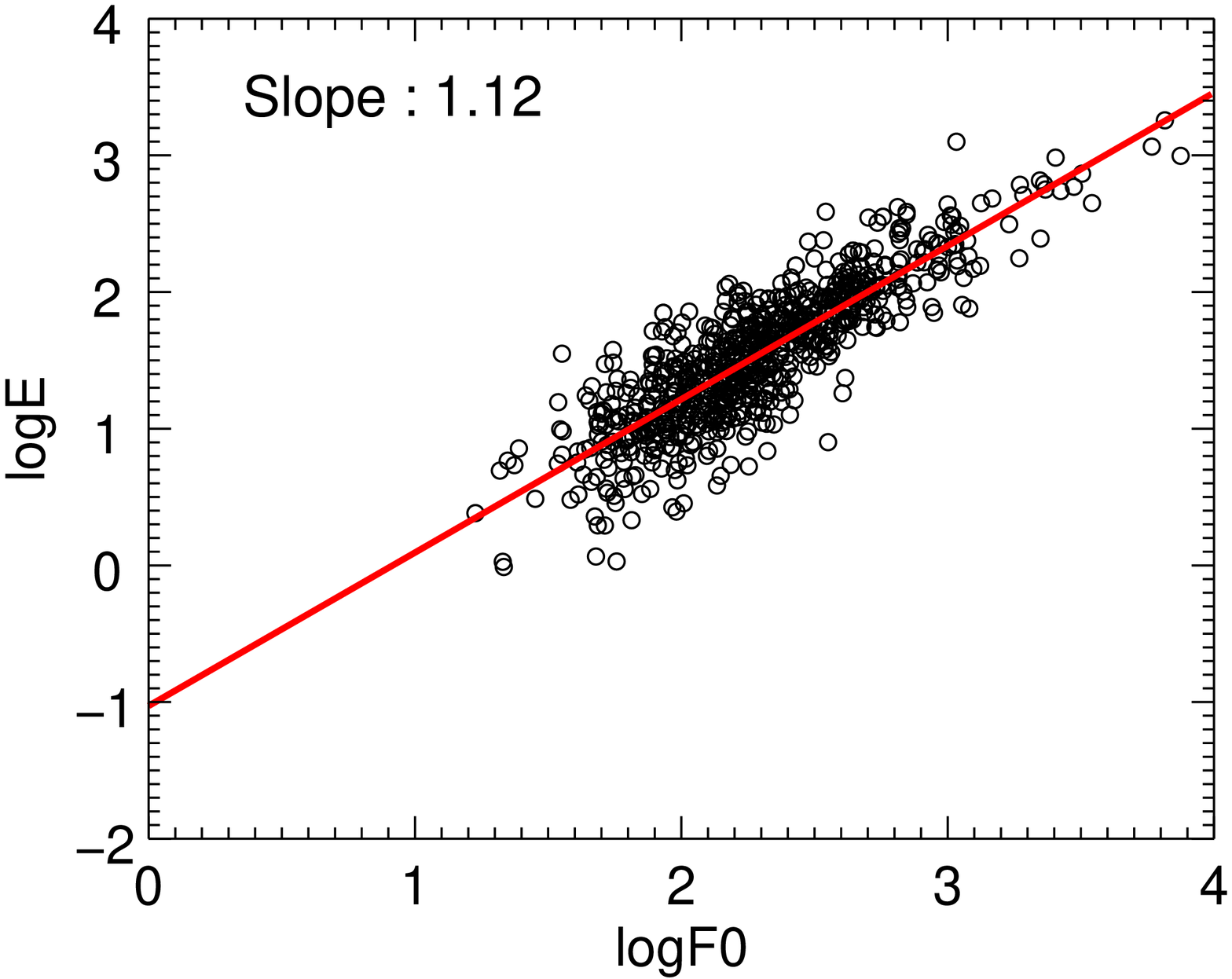}
   \end{minipage}\vspace{0.001cm}
   \caption{Distributions of log~$E$ against log~$D$ (left), log~($T_r/T_d$) (middle) and log~$F_0$ (right).}
   \label{fig:energy}
 \end{figure*}

 To further investigate the properties of blazar flares we calculated the ``energy'' of each flare in the light curve of W2R 1926$+$42 by integrating the flux of each flare over its duration. The modest positive correlation between the values of log\ $D$ and log\ $E$ is shown in the left of Figure \ref{fig:energy}, where the best-fit linear slope is 1.36, with $R^2 = 0.47$. This is no surprise, in that any correlation between length and energy is expected to be positive unless the longer flares are, on average, of substantially lower amplitude.

 A plot of log$E$ against log($T_r/T_d$)  is shown in the middle of Figure \ref{fig:energy} and no correlation is seen between the skewnesses of flares and their total energies. In the right of Figure \ref{fig:energy}, displays the positive correlation between log$F_{0}$ and log$E$ with a best fitting linear slope of 1.12. Again, a positive correlation is expected given the definition of $E$.

 The results of all these tests are summarized in Table~\ref{tab:logpara}.

\subsection{Comparison using SAP Flux data}

\begin{figure*}
    \centering
    \includegraphics[trim=0.2cm 3cm 0.2cm 3cm,width=0.8\textwidth,clip]{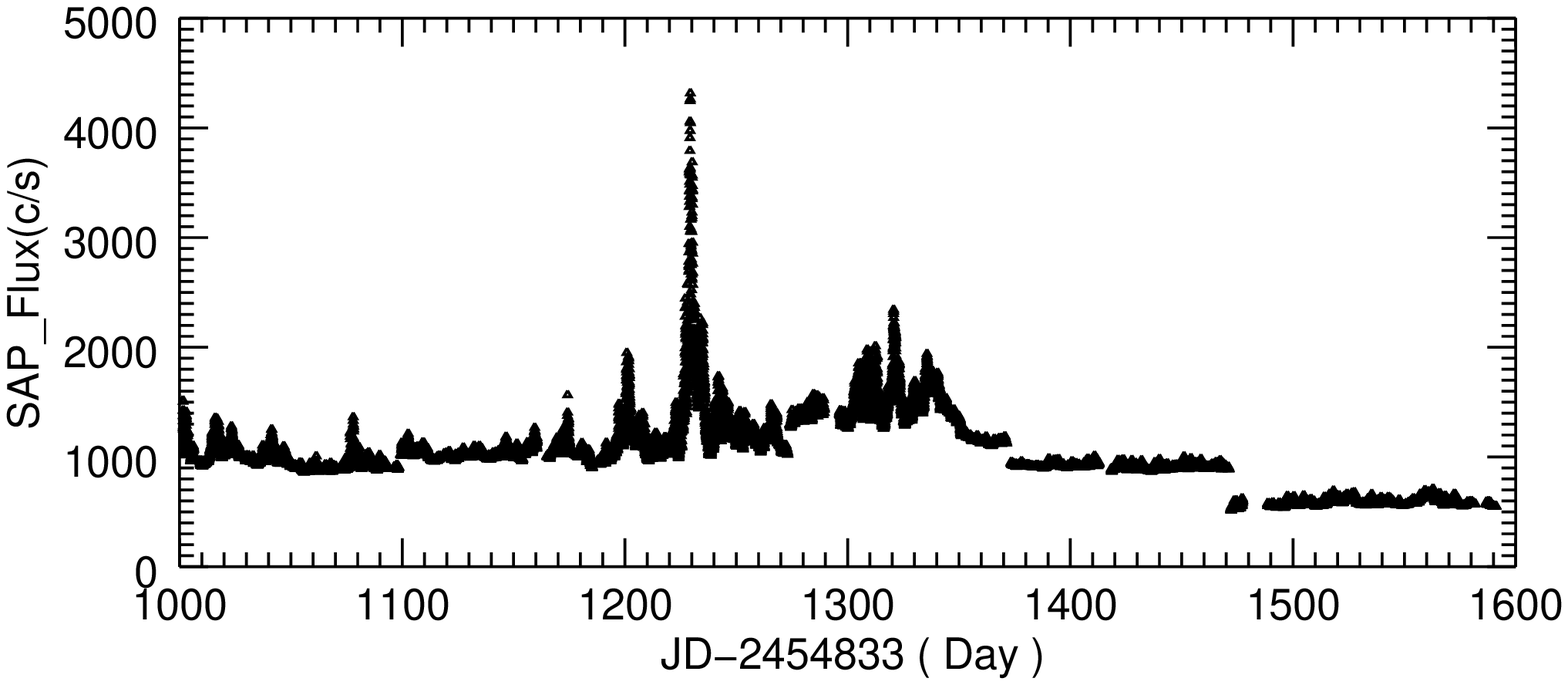}
    \caption{The SAP light curve of W2R 1926+42 from Kepler.}
    \label{fig:SAPlightcurve}
 \end{figure*}

 In case that the corrected flux time series did remove real astrophysical variability in stochastically variable sources such as blazars, we also analyzed the properties using the raw SAP Flux data. The light curve of W2R 1926+42 using SAP Flux data is presented in Figure \ref{fig:SAPlightcurve}.  The nominal count rates are slightly lower for the SAP data but they otherwise are generally very similar.
 The SAP fluxes, do however, show two distinct jumps in the last two quarters, as well as more modest ones between the other quarters.  These are clearly non-physical and are dominated by the different sensitivities of the different modules on which the flux from the source falls but other systematic error sources from the telescope and spacecraft, such as pointing drift, focus changes, and thermal transients, are also present.

 Using the same methods as for the PDC Flux data, we analyzed the SAP Flux variability. Again, we subtracted off a long-term baseline component using polynomial functions and fit the flares in separate quarters so each quarter is taken as independent.
 Adopting the SAP Flux data we find that the mean of the skewness remains at 0.05 and the percentages of the skewnesses are:  $|K| < 0.3$, $35.4\%$; $K < - 0.3$, $30.7\%$; and $K > 0.3$, $33.9\%$.
 It is clear that both versions of the Flux data (PDC and SAP) indicate that most flares are asymmetric.
 We also found similar log-normal and Gaussian distributions for the durations ($D$) and the amplitudes ($F_{0}$) of flares employing the SAP Flux variability.
 The modest positive correlation between the values of log\ $D$ and log\ $E$ shows that the best-fit linear slope is 1.44, with $R^2 = 0.48$,
 and the positive correlation between log$F_{0}$ and log$E$ has a best fitting linear slope of 1.15 ($R^2 = 0.76$).
 Therefore, there does not appear to be any significant differences between these two versions of the blazar light curve when analysed in this way.

\subsection{Lomb-Scargle periodogram}

 The Lomb-Scargle periodogram \citep{Lomb76,Scar82} is a method to estimate the power spectral density (PSD) generated by a physical process manifested through the unevenly sampled time series data.
 The periodogram evaluated from a light curve $a(t_{k})$ spanning n points is determined by a least squares fit to a mean subtracted time series using
 \begin{equation}\nonumber
  a(t_{k}) = C_{1}{\rm sin}(2\pi f_{j}(t_{k}-\omega)) + C_{2}{\rm cos}(2\pi f_{j}(t_{k}-\omega))
 \end{equation}
 and is given by \cite{Horne86}
 \[
  P(f_{j}) =\frac{1}{2\sigma^{2}}\Bigl[\frac{(\sum\nolimits_{i=1}^n{(a(t_{k})-\bar{a})\cos(2\pi f_{j} (t_{k}-\epsilon)})^{2}}{\sum\nolimits_{i=1}^n{\cos^{2} 2\pi f_{j}(t_{k}-\epsilon)}}
 \]
 \begin{equation}
 \ \ \ \ \ \ \ \ \ \ \ + \frac{(\sum\nolimits_{i=1}^n{(a(t_{k})-\bar{a})\sin 2\pi f_{j} (t_{k}-\epsilon)})^{2}}{\sum\nolimits_{i=1}^n{\sin^{2} 2\pi f_{j}(t_{k}-\epsilon)}}\Bigr]
 \end{equation}
 with $\sigma^{2}$ being the variance and $\bar{a}$ the mean value of the time-series $a(t_{k})$. The time shift parameter $\epsilon$ is given by
  \begin{equation}
    \tan(4\pi f_{j}\epsilon) = \frac{\sum\nolimits_{i=1}^n{\sin(4\pi f_{j}t_{k})}}{\sum\nolimits_{i=1}^n{\cos(4\pi f_{j}t_{k})}}
  \end{equation}
The periodogram is evaluated using the algorithm presented in \cite{Press89} in order to achieve fast computational speeds.
We used power law model to constrain the PSD shape, given by
 \begin{equation}
  I(f_{j})=Af_j^\alpha + C,
 \end{equation}
 with amplitude $A$, slope $\alpha$, and a constant Poisson noise $C$.

 We used two methods to compute the power law parameters, including or not including Hanning window functions.
 The Hanning function is used to create a ``window" for Fourier Transform filtering \citep{Har78,Mart14}. It is defined as
  \begin{equation}
   w(k) = 1/2 - 1/2*\cos(2 \pi k/ N), \ \ k = 0,1,...,N-1
  \end{equation}
 Because of the factor of $1/N$ (rather than $1/(N-1)$) in the above equation, the Hanning filter is not exactly symmetric, and does not go to zero at the last point. The factor of $1/N$ is chosen to give the best behavior for spectral estimation of discrete data.
 Hanning windows are expected to perform well in large part due to having features that both include smoothness, which helps to enforce adiabaticity, and composite pulse structure that allows cancellation of errors between different (e.g., diabatic) parts of the pulse \cite{Theis16}.
 The spectral parameters of power law, log parabolic and their $\chi^2$ values are presented in Table~\ref{tab:psd} for each quarter. The median slopes of these individual PSD power laws are $-1.27$ and $-1.44$ for without any window function and including the Hanning windowing, respectively. The results show that the index of power law are usually quite similar with or without the inclusion of the Hanning window.
  A similar conclusion was reached in the analysis of this source's light curve in \cite{Edel13,Mohan16}, though those analyses produced somewhat steeper slopes of the indices computed using power law PSD models.  A PSD slope of $-1.8$ was reported in \cite{Edel13}, and a weighted mean PSD slope of $-1.5 \pm 0.2$ was found in \cite{Mohan16}.

  \begin{table}
    \centering
    \caption{ Results from the parametric power law PSD model fit to the periodogram .}
    \label{tab:psd}
    \begin{tabular}{ccccc}
    \hline\hline
     Quarter & Window function & log(A) & $\alpha$ & $\chi^2$ \\
    \hline
       Q11        & no      & 0.11  & -1.27 & 2.28 \\
    (1009-1094d)  & hanning & -0.45 & -1.36 & 6.71 \\
       Q12        & no      & 0.13  & -1.23 & 2.10\\
    (1099-1182d)  & hanning & -0.40 & -1.37 & 5.67\\
       Q13        & no      & -0.01 & -1.44 & 3.26\\
    (1182-1273d)  & hanning & -0.25 & -1.43 & 5.51\\
       Q14        & no      & -0.13 & -1.45 & 3.12\\
    (1274-1371d)  & hanning & -0.42 & -1.44 & 5.48 \\
       Q15        & no      & 0.17  & -1.12 & 2.77\\
    (1373-1471d)  & hanning & -0.46 & -1.46 & 6.53\\
       Q16        & no      & 0.13  & -1.23 & 1.85 \\
    (1472-1558d)  & hanning & -0.39 & -1.50 & 5.49\\
       Q17        & no      & -0.03 & -1.42 & 3.12\\
    (1559-1582d)  & hanning & -0.35 & -1.51 & 5.76\\
    \hline
    \end{tabular}
  \end{table}

\section{Discussion}

 Detailed and systematic investigation of blazar variability provides crucial information on the location and the structure of the energy emission zones in blazar jets. It also provides insights into the radiative and particle acceleration mechanisms responsible for the production of the observed emission. The {\it Kepler} satellite provides extraordinarily  well sampled data over very extended periods for a limited number of blazars and thus allows us to examine flare characteristics in a more sophisticated fashion than was possible previously.

 The dominant source of strong variability from blazar jets on longer timescales is understood to be the interaction of a shock wave with the jet plasma \citep{Bland79,Mars85}. Short timescale variability can reflect physical processes in smaller emitting regions of a jet without any direct relation to more slowly varying components \citep{Sasada17}.  Light curves of blazar flares in both X-rays and $\gamma$-rays have shown similar rise and decay timescales \citep{Abdo10b,Chat12,Wie16}. Symmetric shapes of the light curves can strongly constrain the injection and cooling timescales \citep{Chiab99}. Assuming that the electrons are distributed in energy as a power-law function, \cite{Chiab99} showed symmetric light curves are expected when the cooling time of electrons $t_{cooling}$ was much shorter than light crossing time, $R/c$, with $R$  the size of emission region. However, when $t_{cooling} > R/c$, the timescale for decline is longer than rise timescale, leading to asymmetric profiles.

 The range of the Doppler factor, $\delta$, of BL Lacertae objects in radio and optical wavelengths is between $1.1-24.0$ \citep{Hov09,Fan16}, and the peak value of the Doppler factor distribution is around 5. \cite{Hu17} studied the physical properties of the high-energy emission region by modelling the quasi-simultaneous multi-wavelength (MWL) spectral energy distributions of 27 Fermi-LAT detected low-synchrotron-peaked (LSP) blazars, and showed that the derived physical parameters indicated Doppler factors ranging from 7 to 36. The duration  $D_j$ in the jet frame is related to the observed $D$ by $D_{j}=\delta D /(1+z)$, corrected for the cosmological redshift $z$. Taking the typical value $\delta = 10$, $D_j$ ranges between $ 3.8\times 10^{4}$\ s and  $2.2\times 10^{6}$\ s \ (0.4 -- 26.0 \  d). Therefore, the size of the emission region $R \ (R=cD_{j})$ can be estimated as $1.1\times 10^{15}$ -- $6.6\times 10^{16}$~cm.

 The jet frame synchrotron cooling  timescale can be expressed as a function of electron energy and the magnetic field \citep{Chen11}
  \begin{equation}
   \centering
   \tau_{cooling} {\rm(s)} =\frac{7.7 \times 10^{8}}{\gamma B^{2}}.
   \label{equ:coolingtime}
  \end{equation}

  \begin{table}
    \centering
    \caption{Estimated flare parameters}
    \label{tab:parameter}
    \begin{tabular}{lcc}
    \hline\hline
     Parameter  & Range \\
    \hline
       observed duration $D$ (s)         & $4.5\times 10^{3}$ -- $2.6\times 10^{5}$   \\
       median duration  $D_{m}$ (s)      & $4.7\times 10^{4}$                          \\
      restframe duration $D_{j}$(s)      & $3.8\times 10^{4}$ -- $2.2\times 10^{6}$    \\
       emission region size $R$(cm)      & $1.1\times 10^{15}$  -- $6.6\times 10^{16}$ \\
       $\tau_{cooling}$ (s)              & $1.7 \times 10^{4}$  -- $8.6 \times 10^{6}$ \\

    \hline
    \end{tabular}
  \end{table}

 \citet{Mars14} constrained the magnetic field strength and electron Lorentz factors to $0.01 < B < 0.5$ G and $\gamma_{max} \geq 10^{4}$ for blazars. Similar values of magnetic field strengths ranging between $0.3-2.0$ G were derived by \cite{Hu17} using 27 LSP blazars, which are smaller than the results found by \cite{Ghis10}, but larger than the results of \cite{Yan12}. \cite{Ghis10} found that there were somewhat larger magnetic fields in FSRQs (1-10 G) than in BL Lacs (0.1-1 G). However, the distribution of $B$ shown in \cite{Yan12} was in a narrow range of $0.15-0.25$ G. The value of the electron energy was estimated by \citet{Zhang16} with $\gamma_{min}= 10^{3}, \gamma_{max}= 5 \times 10^{4}$ in the blazars' emission regions. Given the above estimates, here we adopt $B=0.3$ G, $\gamma_{min}= 10^{3}$, and $\gamma_{max}= 5 \times 10^{4}$  to estimate the cooling timescales of W2R 1926+42. Therefore, according to Equation~\ref{equ:coolingtime},  $1.7 \times 10^{4} $ s $< \tau_{cooling} <8.6 \times 10^{6}$ s. These parameters are summarized in Table~\ref{tab:parameter}.

 The interpretation that the flare evolution is governed by the geometry of the active region seems to be supported by \cite{Chiab99} and \cite{Kat00}. However, symmetry can be observed only if all other (energy dependent) timescales are shorter than the blob crossing time and any other relevant geometrical timescales. As for the light crossing time effects within the active region, there are two main aspects \citep{Chen11}. The first is a purely geometric effect, called ``external'' \citep{Katar08}, while the second, ``internal'',  effect includes the impact of actual physical evolution of the variability and is the real challenge for multi-zone modeling. \cite{Chen11} presented three scenarios, aimed at modeling the variability exhibited by Mrk 421 during the 2011 March 19 flares. However, the symmetric flare shapes could be produced only by scenarios where the blob is filled with a pre-existing (background) electron population, homogeneous throughout the volume. These electrons served as a slowly evolving component in the electron distribution, and participated fully in the time evolution of the blob, as they cooled and emitted radiation. When photons emitted by electrons (and seed photons) vary on  timescales shorter than the geometric timescales, the latter one will dominate the flare profile, hence a symmetric flare will be produced. Conversely, when the emission processes timescales are longer than the geometric ones, a slower cooling decay profile emerges.

  According to the HMFM model proposed by \cite{Zhang14} to analyze the asymmetric variability of flares,
  the light curves generally will not be symmetric in time because of the asymmetry in timescales between the dynamics of the shock moving through the emission region and the electron cooling. However, this model can produce symmetric light curves if the size of the active region is much smaller than the overall jet emission region.

  We note from Table ~\ref{tab:parameter} that there is substantial overlap between the estimated light travel times and the allowed range of cooling times, $\tau_{cooling}$. The profiles of flares are mostly asymmetric with  $66.0 \%$ evincing $|K |> 0.3$. Combined with the parameters we estimated above, it does not appear that the interpretation of asymmetric light curves suggested by \cite{Chiab99} can be applied to explain our results. Perhaps the phenomenon that the flares of W2R 1926+42 are predominantly asymmetric can be explained using HMFM model. In that case the times for the shock to move through the emission region and the electron cooling are substantially different. The distributions of log-normal statistics for flare duration and flux variance also may be generated within a jet-in-jet statistical model \citep{Gia09,Bit12}, suggesting that the variability stems from multiplicative processes.

\section{Conclusions}

 We have studied some of the time variability properties of  the blazar W2R 1926+42 in data obtained by {\it Kepler}. The light curves (up to 10 days segments) were investigated systematically. Adopting the form of exponential rise and decay for individual flares \citep{Val99, Abdo10b,Chat12}, we decomposed the light curve into often overlapping flares. The skewness of the individual flares was studied and we found that the majority of flares are asymmetric, with $66.0 \%$ having $|K |> 0.3$. The light travel time estimates for the size of the emitting regions, $D_{j}$, is between $\sim 4 \times 10^{4}$ s and  $\sim 2 \times 10^{6}$s. These emission regions  have estimated physical sizes in the range $\sim 1\times 10^{15}$ cm to $\sim 7 \times 10^{16}$ cm.

 We studied the observed distributions of duration, $D$, and peak flux, log$F_{0}$, for each flare. The adjusted R-square test results showed that $D$ and log$F_{0}$ could be well fitted by log-normal and Gaussian distributions, respectively. The log-normal distribution could be interpreted in terms of a jets-in-a-jet model, as  a log-normal distribution may be produced by magnetic dissipation triggering the emission of extremely relativistic blobs of plasma \citep{Gia09,Bit12}.

 The correlation between log~$E$  and log~$D$  has a more significant positive slope of 1.36, indicating that the longer the duration of a flare is, the larger its energy is. There is a similar positive correlation between log~$E$ and log~$F_{0}$ (slope of 1.12). Both of these are expected to show positive correlations through our definition of the ``energy'' as the integral of the flux over the duration, but modeling the strengths of these correlations is beyond the scope of this paper and will be attempted in future work.

 We also examined the shape of the PSD shapes and found good fits to power-law shapes, with slopes ranging between $-1.1$ and $-1.5$ for the seven quarters of {\it Kepler} observations.  The median slopes of these power law were $-1.27$ without any window function applied and $-1.44$ when Hanning windowing was used.

\section*{Acknowledgements}
 We are grateful to the referee for insightful comments and constructive suggestions that have helped us to improve the paper.
 This work is supported by the National Natural Science Foundation of China under grants Nos.\ 11203016 and 11143012. This work is partly supported by Natural Science Foundation of Shandong province (Nos. JQ201702, ZR2017PA009) and Young Scholars Program of Shandong University, Weihai. Yutong Li is grateful for support from the China Scholarship Council. PJW is grateful for hospitality at KIPAC, Stanford University, and the Shanghai Astronomical Observatory during sabbatical visits. ACG is partially supported by the Chinese Academy of Sciences President's International Fellowship Initiative (PIFI) (grant No.\ 2016VMB073). This paper includes data collected by the Kepler mission. Funding for the Kepler mission is provided by the NASA Science Mission directorate. The authors would like to thank the Kepler mission team for producing this extraordinary data set.




\bsp	
\label{lastpage}
\end{document}